\documentclass{article}

\usepackage[pdftex]{graphicx}
\usepackage[cmex10]{amsmath}

\usepackage{fullpage}  
\usepackage{paralist}

\def\A{\mathbf{A}}
\def\G{\mathbf{G}}

\def\R{\mathbf{R}}

\def\W{\mathbf{W}}

\DeclareMathOperator{\E}{E} 
\DeclareMathOperator{\cov}{Cov}

\def\adj{^\dagger }
\def\inv{^{-1}}
\def\diag{\mathrm{diag}}

  \def\sautpage{ }

\begin{document}

\title{ Component separation with flexible models.\\ Application to the separation of astrophysical emissions}

\author{
  Jean-Fran\c cois Cardoso${}^{1,2}$,
  Maude Martin$^2$,
  Jacques Delabrouille$^2$,\\
  Marc Betoule$^2$,
  Guillaume Patanchon$^2$
}

\date{
  \small
  1: Laboratoire de traitement et communication de l'information (LTCI),\\
  CNRS : UMR 5141 TELECOM ParisTech,  46, rue Barrault, 75634 Paris Cedex\\
  2: AstroParticule et Cosmologie (APC),
  CNRS : UMR 7164 - Universit\'e Denis Diderot - Paris VII
}

\maketitle

\begin{abstract}

  This paper offers a new point of view on component separation, based on a model of
  additive components which enjoys a much greater flexibility than more traditional linear
  component models.  
  This flexibility is needed to process the complex full-sky observations expected from
  the Planck space mission, for which it was developed, but it may also be useful in any
  context where accurate component separation is needed.

\end{abstract}

\section{Introduction}

This paper describes an advanced component separation technique and its application to the
analysis of multi-spectral observations of the Cosmic Microwave Background (\textsc{CMB}).
It is developed more specifically in view of the data processing pipeline for the
forthcoming Planck mission~\cite{plancksite} of the European Spatial Agency (\textsc{ESA}).

Component separation is a critical part of \textsc{CMB} data processing because no
frequency channel exists that would be sensitive only to \textsc{CMB} emission.
As an illustration, figure~\ref{fig:inputs} shows (simulated) maps of the microwave
emission of the sky observed in the frequency channels of Planck \textit{i.e.} in spectral
bands centered around frequencies $\nu= 30, 44, 70, 100, 143, 217, 353, 545,
857~\mathrm{GHz}$.  The red (in an arbitrary color map) horizontal strips along the
Galactic equator, in the middle of the maps, are due to Galactic emissions (that is,
emissions from \emph{our} Galaxy).  In contrast, towards the poles where the Galactic
emissions are weaker, one can see in the center channels (100, 143 and 217~GHz), an
homogeneous texture which is the signal of interest: the (spatial) fluctuations (or
`anisotropies') of the~\textsc{CMB}.
Unfortunately, even in the polar areas, the contamination of the \textsc{CMB} by various
astrophysical emissions (or 'foregrounds', as opposed to the cosmic \emph{background}) is
still significant.  There, especially at the smallest angular scales, contamination is
dominated by the emission of remote extragalactic objects -- galaxies (or `point sources')
of various types (which emit radiation through similar physical processes as our own
Galaxy), and galaxy clusters (which emit via the Sunyaev-Zel'dovich effect, hereafter: the
`SZ component').

There is no significant occlusion between the various physical components so that the
signal $X_\nu(\xi)$ measured at a frequency $\nu$ in direction $\xi$ is a linear
superposition:
\begin{displaymath}
  X_\nu(\xi) = X_\nu^\mathrm{cmb}(\xi) + X_\nu^\mathrm{gal}(\xi) + X_\nu^\mathrm{SZ}(\xi)+ \ldots + N_\nu(\xi)
\end{displaymath}
where $X_\nu^\mathrm{x}(\xi)$ is the contribution due to a particular component
$\mathrm{x}$ and $N_\nu(\xi)$ accounts for instrumental noise at the map level.

It must be stressed that, even though the lowest (30, 44, 70~GHz) and highest (353, 545,
857~GHz) frequency channels are more contaminated by Galactic emissions, they still are
very informative because they can be used to predict and remove the Galactic emission in
the center channels, leaving only \textsc{CMB} and noise\ldots\ in an ideal data
processing scenario.
Successful scientific exploitation of the \textsc{CMB} measurements for cosmological
studies critically depends on accurate cleaning of the \textsc{CMB} and on quantifying as
well as possible any remaining contamination.  Figure~\ref{fig:reccmb} illustrates
on simulated data what can be achieved with our technique on a very large fraction of
the sky.

Even though \textsc{CMB} recovery is the main goal, recovering the other components also
is of scientific interest.  For instance, Planck should also yield a very rich catalog of
galaxy clusters, and provide a lot of information about the emission of the interstellar
medium in our own Galaxy.  Hence, we are actually facing a problem of \emph{component
  separation} from multiple observations.

The component separation problem is not new to the astrophysics community.  In the past
ten years however, component separation for CMB observations specifically has motivated
the development of a large variety of methods with various degrees of sophistication (see
\cite{LNP-Valencia} for a review).  The most direct methods rely on modelling and
subtracting foreground emission. Such methods, typically, involve a thorough understanding
of the emission processes, necessary to build the foreground model.  Other methods use
simple decorrelation, or make use of differences between properly weighted maps
\cite{2006ApJ...648..784H}. The processing of data from the WMAP satellite mission makes
uses of both kinds of methods \cite{2003ApJS..148...97B}.

The large majority of existing component separation methods in the context of CMB
observations rely on a `mixing model' as follows: for any available direction $\xi$ in the
sky, the $m$ channels at frequencies $\nu_1,\ldots, \nu_m$ provide $m$ data points which
can be collected in an $m\times 1$ vector $X(\xi)$ which is modelled as
\begin{equation}\label{eq:multmod}
  X(\xi) = \A S(\xi) +N(\xi)
\end{equation}
where each entry of $S(\xi)$ contains the sky-pattern for the emission of a given
component.  The emissivity of a given component depends on the frequency $\nu$; hence,
each column of matrix $\A$ reflects the emission law of the corresponding component.
This equation, then, forms the basis for many component separation algorithms but it forces a
special structure on each component, namely that it `scales rigidly with frequency'. Some methods
are developed to solve this component separation problem when the mixing matrix $A$ is
known {\emph{a priori}}. This is the case, for instance, of the work described in
\cite{1996MNRAS.281.1297T,1998MNRAS.300....1H,1999NewA....4..443B,2002MNRAS.336...97S}.

Component separation is a well studied problem in Signal Processing, but the nature of the
available data and the scientific objectives call for specific techniques.
In particular, one may think of resorting to Independent Component Analysis (\textsc{ICA})
methods since these `blind' techniques may be useful to deal with some of the
uncertainties in the data structure.  In addition, columns of the mixing matrix $\A$ corresponding to
some of the astrophysical components (e.g. components modelling the emission of our own Galaxy)
are known a priori up to significant uncertainty, and more precise knowledge is an objective of the
forementioned astrophysical observations. Implementations of ICA ideas in the context of CMB 
observations can be found in 
\cite{2002MNRAS.334...53M,2002AIPC..617..125S,2003MNRAS.346.1089D,2003MNRAS.345.1101M,2006MNRAS.373..271B}.

Compared with all the above methods, our
framework allows the inclusion of components with arbitrary structure (at least in terms of
second order correlations). This structure, as will be detailed later on, can be matched to 
prior knowledge of the nature of the astrophysical emission. 
Along the same spirit of matching a parametric model of the emissions to the data, although quite
different in implementation, are the work described in
\cite{2000ApJ...530..133T,2006NewAR..50..861E}.

Although there is substantial motivation for letting the data alone indicate the
appropriate model of the emissions, a fully blind approach would leave out a lot of
valuable prior information in the CMB context.  The approach described in this paper is
rooted in \textsc{ICA} but allows for the inclusion an arbitrary amount of prior
information, in particular by allowing components with an arbitrary structure.

Again, the approach to component separation described herein is not specific to this
particular problem and may be considered for application to any situation where
`expensive' data deserve special care and have to be fitted by a complex component model.

The paper is organized as follows: 
section~\ref{sec:model} describes our approach, 
section~\ref{sec:discussion} focuses in specific aspects; 
implementation is discussed in section~\ref{sec:impl} in terms of a library of
components; 
section~\ref{sec:results-simul-planck} illustrates an application of the method on data
from the Planck sky model.

\sautpage

\section{A flexible approach to component separation}\label{sec:model}

This section introduces a key ingredient of our approach: the development of a flexible
model of \emph{additive components} in which the notion of `mixing matrix', which is
central to standard \textsc{ICA}, is disposed of.  Not only the mixing matrix is not
needed but it is `considered harmful'.

\subsection{Unconstrained components}

We start the exposition by considering the most general component model for an $m\times 1$
random vector $X$: we only postulate that $X$ is made of the linear superposition of $C$
components:
\begin{equation}\label{eq:compmodel}
  X = \sum_{c=1}^C X^c
\end{equation}
where each component $X^c$ is an $m\times 1$ random vector.  For simplicity, we shall
assume that all components have zero mean.  
Denote $\R$ (\textit{resp.}  $\R^c$) the covariance matrix of vector $X$
(\textit{resp.}  of its $c$-th component $X^c$).  
A key assumption of our model is mutual decorrelation between all components, which implies
that the covariance matrix of $X$ decomposes as
\begin{equation}\label{eq:deccov}
  \R = \sum_{c=1}^C \R^c .
\end{equation}
Consider now recovering linearly a particular component $X^c$ from the
superposition~(\ref{eq:compmodel}).  Denote $\W_c$ the $m\times m$ matrix which best
predicts $X^c$ based on $X$ in the sense that
\begin{equation}\label{eq:defWc}
  \W_c=\arg\min_\W \E|\W X - X^c|^2.  
\end{equation}
this problem is easily solved under our assumptions and one readily finds:
\begin{equation}\label{eq:wiener}
  \W_c = \R^c \R\inv
\end{equation}
At this stage, we wish to deliver a simple but important message: eq.~(\ref{eq:wiener})
means that component separation, understood as computing $\widehat X^c = \W^c X$, requires
only knowing $\R^c$ and $\R$.  Of course, covariance matrix $\R$ may (but need not to) be
simply estimated from the empirical covariance matrix $\widehat \R$ of $X$.  Therefore,
component separation is solved if we can uniquely solve the \emph{covariance separation
  problem}, in the sense of identifying the component terms $\R^c$ in
decomposition~(\ref{eq:deccov}).

In order to achieve covariance separation, some additional information or assumptions are
of course needed.  Our method jointly exploits two possibilities: considering
\emph{localized matrices} (for more information) and considering \emph{structured
  matrices} (for more constraints).

\subsection{Localized statistics, localized separation}\label{sec:Loc}

This section defines the `localized statistics' on which our method is based.  
Consider the component separation problem based on $p$ samples on $m$ detectors.  We shall
not compress this data set into a single sample $m\times m$ covariance matrix but rather
`localize' the statistics and form a \textit{set} of \emph{localized covariance matrices}.
By `localization', we typically mean localization in time, space, frequency, wavelet
space, depending on the problem at hand.
The simplest example of localization is suggested by the separation of non-stationary time
series as considered in \cite{ParraSpence:ieee,nstatIEEESP} where the data are
time-indexed and one would divide the observation interval into $Q$ sub-intervals and
compute an estimate of the covariance of $X$ over each of these intervals.

More generally, we assume that the data are available as $p$ vectors $X(1),\ldots, X(p)$,
each of size $m\times 1$ where the $i$-th entry of $X(j)$ is either the $j$-th sample of
the $i$-th input signal or its $j$-th coefficient in some basis (\textit{e.g.}  the
Fourier basis).  In our application to \textsc{CMB} analysis, we use the spherical
harmonic basis (see section~\ref{sec:results-simul-planck}).
Basis choice is discussed below; it does have an impact on separation but it does not
change the nature of the problem in the sense that, if the component
model~(\ref{eq:compmodel}) holds in some basis, it also holds in any other basis and that
if a component is reconstructed in a given basis, it becomes available in any basis.

Localized statistics are formed by dividing the index set $[1,\ldots, p]$ into $Q$ subsets
which are called `domains' in the following (time domains would be time intervals,
spectral domains would be frequency bands, etc\ldots).
Denoting $\mathcal{D}_q$ the $q$-th domain, partitioning of the data set reads $[1,\ldots,
p ] = \cup_{q=1}^Q \mathcal{D}_q$ and localized statistics are computed (defined) as
\begin{equation}\label{eq:empscm}
  \widehat\mathcal{R}= \{\widehat\R_q\}_{q=1}^Q ,
  \qquad
  \widehat\R_q = \frac 1{p_q} \sum_{j\in\mathcal{D}_q}  X(j)X(j)\adj ,
\end{equation}
where $p_q$ is the number of coefficients in domain $\mathcal{D}_q$.

The expected value of the sample covariance matrix for the $q$th domain is defined/denoted
as
\begin{displaymath}
  \R_q =  \E \widehat\R_q
\end{displaymath}
and we shall use the same notation for each component so that, these being mutually
uncorrelated by assumption, one has the decomposition
\begin{equation}\label{eq:decRq}
  \R_q = \sum_{c=1}^C \R_q^c 
  \qquad
  \R_q^c 
  =  \E
  \Bigl\{
    \frac 1{p_q} \sum_{j\in\mathcal{D}_q}  X^c(j)X^c(j)\adj 
  \Bigr\}
  =
  \frac 1{p_q} \sum_{j\in\mathcal{D}_q}  \cov X^c(j)
\end{equation}
We can then perform a localized separation in the sense that the localized Wiener filter
for recovering component $c$ in domain $q$ is
\begin{equation}\label{eq:locwiener}
  \widehat X^c(j) = \R_q^c \R_q\inv X(j)
  \ \ \text{for}\ \ j\in\mathcal{D}_q 
\end{equation}
A key point is that, by definition, this filter is specialized to operate on domain $q$,
hence it is adapted to the local correlation conditions.

\subsection{Constrained components and model identification}\label{sec:constr-comp-model}

Implementing component separation by (\ref{eq:locwiener}) requires knowning the local
covariance matrices $\R_q^c$.  These are usually unknown and must be estimated, based on
$\widehat\R_q$, the sample estimate~(\ref{eq:empscm}) of their sum.
It goes without saying that, without constraining the possible values of $\R_q^c$, it is
not possible in general to uniquely resolve (an estimate of) this sum into its components.
Constraining a given component $c$ is achieved by parametrizing the matrix set
$\mathcal{R}^c=\{\R_q^c\}_{q=1}^Q$ by a parameter vector $\theta^c$ \textit{i.e.}  a
parametric model for the $c$-th component is a function $\theta^c \rightarrow
\mathcal{R}^c(\theta^c)=\{\R_q^c(\theta^c)\}_{q=1}^Q$.
Some examples of these parametric models are given in section~\ref{sec:mylib}.

A parametric model $\theta\rightarrow\mathcal{R}(\theta)$ follows by assuming component
decorrelation~(\ref{eq:decRq}) and taking the global parameter $\theta$ as the
concatenation of the parameter vectors of each component:
\begin{displaymath}
  \mathcal{R}(\theta) =
  \{\R_q(\theta)\}_{q=1}^Q = \bigl\{ \sum_c\R_q^c(\theta^c)\bigr\}_{q=1}^Q 
  \qquad
  \theta=(\theta^1,\ldots,\theta^c) .
\end{displaymath}


The parametric model is identified by fitting it to data, that is, by minimizing a measure
of mismatch between $\widehat\mathcal{R}$ and $\mathcal{R}(\theta)$ as:
\begin{equation}\label{eq:idcrit}
  \widehat\theta=\arg\min \phi(\theta)
  \quad\text{where}\quad  
  \phi(\theta) = \sum_{q=1}^Q w_q K(\widehat\R_q, \R_q(\theta)) 
\end{equation}
where $K(\cdot,\cdot)$ is measure of mismatch between two positive $m\times m$ matrices
and $w_q$ are positive weights.
Our most common choice is to use
\begin{equation}\label{eq:defWK}
  w_q=p_q 
  \quad \text{and} \quad
  K(\R_1,\R_2)=\frac12 \left[ \mathrm{trace}(\R_1\inv\R_2) -\log\det(\R_1\inv\R_2) -m \right]
\end{equation}
because this is the form which stems from the maximum likelihood principle.  
Specifically, if a coefficient $X(i)$ belonging to domain $q$ is modeled as
$X(i)\sim\mathcal{N}(0,\R_q)$ and as independent from $X(i')$ for $i\neq i'$, then
$\phi(\theta)$ is the negative log-likelihood of the model~\cite{nstatIEEESP}.

\subsection{Summary}

In summary, our method is based on the following steps and design choices:
\begin{compactenum}
\item\label{it:basis} 
  Choice of a basis to obtain coefficients $X(i)$ (original space, Fourier or
  spherical-harmonic space, wavelet space,\ldots).
\item\label{it:domain} 
  Choice of domains $\{\mathcal{D}_q\}_{q=1}^Q$ to localize their second-order statistics
  $\widehat\mathcal{R}$,
\item \label{it:models} 
  Choice of a model $\theta^c\rightarrow \mathcal{R}^c(\theta^c)$, for the contribution
  $\R_q^c$ of each component $c$ to the covariance matrix over each domain $q$.
\item Minimization of the matching criterion~(\ref{eq:idcrit}) to obtain the estimate
  $\hat\theta$, hence estimates $\hat\theta^c$ for the parameters of each component.
\item Estimation of the coefficients of each component by $\widehat X^c(j) =
  \R_q^c(\hat\theta^c) \R_q(\hat\theta)\inv X(j)$, that is the version
  of~(\ref{eq:locwiener}) based on estimated parameters.
\item Reconstruction of the components from their coefficients $\widehat X^c(j)$.
\end{compactenum}
At this stage, most of the statistical framework is in place but our method is not
completely specified yet because of the flexibility in design choices in items
\ref{it:basis}, \ref{it:domain} and~\ref{it:models} above among other issues.  Next
section discusses some of them.

\sautpage
\section{Discussion}\label{sec:discussion}

\subsection{Correlated or multi-dimensional components}

We wish to contrast the standard form $X=\A S+N$ of the noisy linear component model of
eq.~(\ref{eq:multmod}) with the unconstrained model $X=\sum_c X^c$ of
eq.~(\ref{eq:compmodel}).
Obviously, the standard model is included in the unconstrained model since one may always
write $\A S+N =\sum_{c=1}^C X^c$ with $X^c =\mathbf{a}_c S_c$ for $1\leq c\leq n$ and
$X^C=X^{n+1}= N$ where $\mathbf{a}_c$ denotes the $c$-th column of $\A$ and $S_c$ denotes
the $c$th entry of $S$.

A component $X^c(\xi)$ which can be written as $X^c(\xi) =\mathbf{a}_c S_c(\xi)$ is fully
coherent in the sense that any two entries of vector $X^c(\xi)$ are 100\% correlated.  Of
course, this is because all the randomness in $X^c(\xi)=\mathbf{a}_c S_c(\xi)$ comes from
a single scalar random variable $S_c(\xi)$.  A fully coherent component contributes
exactly the same pattern on all sensors (or channels) up to a fixed proportionality
coefficient.
It has a rank-one covariance matrix: $\R^c = \cov(S_c(\xi)) \mathbf{a}_c
\mathbf{a}_c\adj$; for this reason, it could also be called a one-dimensional since it
leaves in the 1D space spanned by vector $\mathbf{a}_c$ or in the 1D range space of
$\R^c$.

In our application, the CMB is a fully coherent component: from one frequency channel to
another, only its overall intensity changes (ignoring the frequency-dependent resolution).
This is in contrast with, for instance, the Galactic emission, which cannot be represented
over the sky as a fully coherent component (see fig.~\ref{fig:decinputs}).

Assume now that, in the model $X=\A S+N$, two sources, say $i$ and $j$, are statistically
\emph{dependent} while all other pairs of sources in $S$ are mutually independent.  These
two sources contribute $\mathbf{a}_i S_i + \mathbf{a}_j S_j$ to $X$ and we decide to lump
them into a single component denoted $X^c$ for some index $c$: $X^c= \mathbf{a}_i S_i +
\mathbf{a}_j S_j$.  Then $X$ can still be written as in eq.~(\ref{eq:compmodel}) and all
the \emph{components} $X^c$ are independent, \emph{again}.  In other words, we represent
here the contributions of two correlated sources as the contribution of a single
2-dimensional component: we now have one less component but it is 2-dimensional.  Clearly,
this component is not fully coherent: it contributes a pattern on sensor $m$ which is not
proportional to the pattern contributed to another sensor $m'$.
One can obviously define multidimensional components of any dimension, each one possible
representing the contribution of several correlated sources.  In section
\ref{sec:results-simul-planck}, we use a 4-dimensional component to model Galactic
emission.

Note that, in eq.~(\ref{eq:compmodel}), the noise term is included as one of the
components.  If the noise is uncorrelated from channel to channel, as is often the case,
then the noise component is $m$-dimensional (the largest possible dimension).

More generally, our model does not require any component to be low dimensional.  Rather,
our model is a plain superposition of $C$ components as in eq.~(\ref{eq:compmodel}).  None
of these components is required to have any special structure, one-dimensional or
otherwise.

\subsection{Library of components}\label{sec:mylib}

We call a collection of parametric models $\theta^c \rightarrow \mathcal{R}^c(\theta^c)$ a
\emph{library} of components.  In practice, each member of the library must not only
specify a parametrization $\theta^c \rightarrow \mathcal{R}^c(\theta^c)$ but also its
gradient and related quantities (see sec.~\ref{sec:comp-deriv}).

Typical examples of component models are now listed.
\begin{compactenum}

\item The `classic' ICA component is fully coherent (one dimensional) $X^c(i) =
  \mathbf{a}_c S_c(i)$.  Denoting $\sigma_{qc}^2$ the average variance of $S_c(i)$ over
  the $q$th domain, the contribution $\R_q^c$ of this component to $\R_q$ is the rank-one
  matrix
  \begin{displaymath}
    \R_q^c = \mathbf{a}_c \mathbf{a}_c\adj \sigma_{qc}^2
  \end{displaymath}
  This component can be described by an $(m+Q)\times 1$ vector $\theta^c$ of parameters
  containing the $m$ entries of $\mathbf{a}_c$ and the $Q$ variance values
  $\sigma_{qc}^2$.  Such a parametrization is redundant, but we leave this issue aside for
  the moment.

\item A $d$-dimensional component can be modeled as
  \begin{displaymath}
    \R_q^c = \A _c P_{qc} \A _c\adj
  \end{displaymath}
  where $\A _c$ is an $m\times d$ matrix and $P_{qc}$ is an $d\times d$ positive matrix
  varying freely over all domains.  This can be parametrized by a vector $\theta^c$ of
  $m\times d + Q \times d(d+1)/2$ scalar parameters (the entries of $\A _c$ and of
  $P_{qc}$).  Again, this is redundant, but we ignore this issue for the time being.

\item Noise component.  A simple noise model is given by
  \begin{displaymath}
    \R_q^c = \mathrm{diag}(\sigma_1^2, \ldots, \sigma_m^2)
  \end{displaymath}
  that is, uncorrelated noise from channel to channel, with the same level in all domains
  but not in all channels.  This component is described by a vector $\theta^c$ of only $m$
  parameters.  A more general model is $\R_q^c = \mathrm{diag}(\sigma_{1q}^2, \ldots,
  \sigma_{mq}^2)$ meaning that the noise changes from domain to domain; then parameter
  vector $\theta^c$ has size $mQ\times 1$.

\item As a final example, for modeling `point sources' in spectral domain, one may use
  $\R_q^c = \R_\star^c$.  Such a component contributes identically in all domains
  corresponding to a flat spectrum.  If, for instance, we assume that this contribution
  $\R_\star^c$ is known, then the parameter vector $\theta^c$ is void.  If $\R_\star^c$ is
  known up to a scale factor, then $\theta^c$ is just a scalar, etc\ldots In the
  demonstration test discussed in \ref{sec:results-simul-planck}, however, we consider
  instead the sum of Galactic and point source emission as one single 4-dimensional
  component -- a choice which the flexibility of the present model allows us to do.

\end{compactenum}

\subsection{About localization}

There are two strong, very different motivations for localizing the statistics.

\textbf{Localization for accuracy.}  The first motivation is separation accuracy.  If the
strength of the various components (including noise) varies significantly across the
domains $\mathcal{D}_q$, reconstruction is improved by the localized filter.
Indeed, the best linear reconstruction of $X^c(j)$ based on $X(j)$ would be, as seen above,
obtained as $\widehat X^c(j)= \cov(X^c(j))\cov(X(j))\inv X(j)$.  This requires knowing
both $\cov(X^c(j))$ and $\cov(X(j))$.  In practice, it seems difficult to obtain estimates
for these matrices for all samples, \textit{i.e.} for each value of $j$.  However, if they
do not vary too much for all values of $j$ across a domain $\mathcal{D}_q$, then a good
approximation to the best reconstruction is~(\ref{eq:locwiener}) \textit{i.e.}  the
reconstruction filter also is localized, taking advantage of the `local SNR conditions' on
domain $\mathcal{D}_q$.

\textbf{Localization for diversity/identifiability.}
Second, the diversity of the statistics of the components over several domains is
precisely what may make this model blindly identifiable.  
For instance, in the basic ICA model (all components are one-dimensional, no noise), if
$X(i)$ are Fourier coefficients and $\mathcal{D}_q$ are spectral bands, it is known that
spectral diversity (no two components have identical spectrum) is a sufficient condition
for blind identifiability.

\subsection{About blindness and the Fisher information matrix}\label{sec:FIM}

Is this a \emph{blind} component separation method?  It all depends on the component
model.  If all components are modeled as `classic' ICA components (see Sec.~\ref{sec:mylib}),
then the method is as blind as regular ICA.  Our approach, however, leaves open the
possibility of tuning the blindness level at will by specifying more or less stringent
models $\theta^c\rightarrow\mathcal{R}^c$ for some or all of the components.

Of course, it may be difficult to predict if a given parametrization ensures the
identifiability of the model: this is to be discussed on a case-by-case basis.  However,
over-parametrization can be tested numerically because the Fisher information matrix (FIM)
is available in our framework.  It has a natural block structure where the block related
to a pair $(c,c')$ of components is the matrix of size $|\theta^c|\times |\theta^{c'}|$:
\begin{equation}\label{eq:cm:defFIMblock}
  [\mathbf{F}(\theta)]_{cc'}
  =
  \frac12 \sum_q p_q\ \mathrm{trace}
  \left(
    \frac{\partial\R_q^ c  (\theta^c  )}{\partial\theta^c   } \R_q\inv(\theta)  
    \frac{\partial\R_q^{c'}(\theta^{c'})}{\partial\theta^{c'}} \R_q\inv(\theta)  
  \right)
\end{equation}
under the assumptions which make the mismatch measure $\phi(\theta)$ proportional to the
log-likelihood (see sec.~\ref{sec:constr-comp-model}).
The FIM is also used for computing (approximate) error bars.

\sautpage
\section{Implementation}\label{sec:impl}

We discuss the practical issue of actually minimizing the mismatch $\phi(\theta)$ using an
arbitrary library of components.

Note that for a noise-free model containing only `classic ICA' components (no other
constraints than being one-dimensional), criterion $\phi(\theta)$ boils down to a joint
diagonalization criterion which can be very efficiently minimized by a specialized
algorithm~\cite{PhamGauss}.  
For a model including only unconstrained multi-dimensional components and noise, it is
possible to use the EM algorithm~\cite{2003MNRAS.346.1089D}.  EM, however, is not
convenient for general component models and, in addition, it appears too slow for our
purposes.
Therefore, we had to consider more efficient and more general optimization procedures.

\subsection{Optimization}

We found the Conjugate Gradient (CG) algorithm well suited for minimizing $\phi(\theta)$.
Its implementation requires computing the gradient $\partial\phi/\partial\theta$ and
(possibly an approximation of) the Hessian $\partial^2\phi/\partial\theta^2$ for
pre-conditioning.
Since $\phi(\theta)$ actually is a negative log-likelihood in disguise, its Hessian can
classically be approximated by $\mathbf{F}(\theta)$, the Fisher information matrix (FIM)
of eq.~(\ref{eq:cm:defFIMblock}).
These computations offer no particular difficulty in theory but we aim at an
implementation it in the framework of a library of components \textit{i.e.}  computations
should be organized in such a way that each component model $\mathcal{R}_c(\theta^c)$
works as a `plug-in'.

\subsection{Computing derivatives}\label{sec:comp-deriv}

\textbf{Computing the gradient.}   The partial derivative of $\phi$ with respect to
$\theta^c$ takes the form
\begin{equation}\label{eq:cm:parderphi}
  \frac{\partial \phi(\theta)}{\partial \theta^c}
  =
  \sum_{q=1}^Q \mathrm{trace}
  \left(
    \G_q(\theta) 
    \frac{\partial\R_q^c(\theta^c)}{\partial\theta^c} 
  \right)
\end{equation}
where matrix $\G_q(\theta) $ is defined as
\begin{equation}\label{eq:cm:defGq}
  \G_q(\theta) 
  = 
  \frac12 w_q \R_q\inv(\theta) \left(\R_q(\theta)-\widehat\R_q\right) \R_q\inv(\theta)  
  .
\end{equation}
Hence the computation of $\partial\phi/\partial\theta$ at a given
point $\theta=(\theta^1,\ldots,\theta^C)$ can be organized as follows.
A first loop through all components computes $\mathcal{R}(\theta)$ by
adding up the contribution $\mathcal{R}^c(\theta^c)$ of each component.
Then, a second loop over all $Q$ domains computes matrices
$\{\G_q(\theta) \}_{q=1}^Q$ which are stored in a common work space.
Finally, a third loop over all components concatenates all partial
gradients $\partial\phi/\partial\theta^c$, each component implementing
the computation of the right hand side of~(\ref{eq:cm:parderphi}) in
the best possible way, using the available matrices
$\{\G_q(\theta)\}_{q=1}^Q$.

\medskip

\textbf{Computing an (approximate) Hessian.}  The Fisher information matrix can be
partitioned component-wise and each block be computed according to
eq.~(\ref{eq:cm:defFIMblock}).
Therefore its computation can be organized with a double nested loop over $c$ and $c'$ as
soon as the code implementing a given component is able to return the matrix set
$\{\frac{\partial\R_q^ c (\theta^c )} {\partial\theta^c}\}_{q=1}^Q$.
A straightforward implementation of this idea may be impractical, though, because this is
a set of $|\theta^c|\times Q$ matrices, possibly very large.
This problem can be alleviated in the frequent case where components have `local'
variables.

\medskip\textbf{Local variables.}
Consider the case when $\theta^c$ can be partitioned into $Q+1$ blocks: $\theta^c = (
\theta_0^c, \theta_1^c, \ldots, \theta_Q^c)$ where, for $q>0$, the sub-vector $\theta_q^c$
affects only $\R_q^c$ while $\theta_0^c$ collects all the remaining parameters
\textit{i.e.}  those which affect the covariance matrix over two or more domains.
We then say that this component model has `local variables'.  This is a fairly frequent
situation which occurs for instance when the power of a component can be freely adjusted
in each domain 
(the simplest example is the `classic' ICA component: $\R_q^c =
\mathbf{a}_c\mathbf{a}_c\adj \sigma_{qc}^2$, for which $\theta_0^c= \mathbf{a}$ and
$\theta_q^c=\sigma_{qc}^2$ for $q=1,\ldots,Q$.
The global parameter vector $\theta$ inherits this structure by being partitioned
accordingly into a `global part' $\theta_0=(\theta_0^1,\ldots,\theta_0^C)$ and $Q$ local
parts $\theta_q=(\theta_q^1, \ldots,\theta_q^C)$.  

A first and major benefit of such a local/global partitioning is that it introduces many
zero blocks in the FIM since then $[\mathbf{F}(\theta)]_{qq'}=\mathbf{0}$ for $1\leq q\neq
q'\leq Q$, allowing the computation of the pre-conditioned gradient
$\mathbf{F}(\theta)\inv \partial\phi/\partial\theta$ to be organized much more
efficiently.

Another benefit of the local/global partitioning is that it makes it easy to implement a
`local optimization': during global optimization over $\theta$, it is possible at any time
to loop through all $Q$ domains and to solve in each domain, possibly exactly, the
sub-problem $ \min_{\theta_q} K(\widehat\R_q, \R_q(\theta_0, \theta_q)\,) $ which is, of
course, of much smaller size than the original problem in most cases.



\subsection{Indeterminations and penalization.}

Examples of section \ref{sec:mylib} show that `natural' component parametrizations often
are redundant.
From a statistical point of view, this is not important: we seek ultimately to identify
$\mathcal{R}^c = \{\R_q^c\}_{q=1}^Q$ as a member of a family described by a mapping
$\theta^c\rightarrow \mathcal{R}^c(\theta^c)$ but this mapping does not need to be
one-to-one.  The simplest example again is for $\R_q^c=\mathbf{a}_c\mathbf{a}_c\adj
\sigma_{qc}^2$ which is invariant if one changes $\mathbf{a}_c$ to $\alpha\mathbf{a}_c$
and $\sigma_{qc}$ to $\alpha\inv\sigma_{qc}$.  
Therefore, there is a 1D set of pairs $(\mathbf{a}_c, \sigma_{qc})$ achieving the global
optimum but they all correspond to a unique value of $\R_q^c$ itself, which is thus
perfectly well defined at the optimum and is the unique quantity needed to reconstruct the
$c$-th component in domain $q$.

The only serious concern about over-parametrization is from the optimization point of
view.  Redundancy makes the $\phi(\theta)$ criterion \emph{flat} in the redundant
directions and it makes the FIM a singular matrix.
Finding non redundant re-parametrizations is a possibility, but it is often simpler to add
a penalty function to $\phi(\theta)$ for any redundantly parametrized component.  
For instance, the scale indetermination of the classic ICA component $\R_q^c =
\mathbf{a}_c\mathbf{a}_c\adj \sigma_{qc}^2$ when parametrized $\theta_0^c = \mathbf{a}_c$
and $\theta_q^c=\sigma_{qc}^2$ ($q>1$) is fixed by adding $\phi^c(\theta^c) =
g(\|\mathbf{a}_c\|^2)$ to $\phi(\theta)$, where $g(u)$ is any reasonable function which
has a single minimum at, say, $u=1$.
Of course, the addition such a term should be reflected in the gradient and the Hessian of
the matching criterion.

\sautpage
\section{Results on simulated Planck data }\label{sec:results-simul-planck}

\def\bA{\mathbf{A}}
\def\bP{\mathbf{P}}
\def\bR{\mathbf{R}}
\def\ba{\mathbf{a}}
\def\bx{\mathbf{x}}

\def\newfigdir{./figs}
\def\psmdir{\newfigdir}
\def\thumbdir{\newfigdir}
\def\dircatania{\newfigdir}

This section describes an application of our framework to a CMB data set.  Although we go
into some detail, several issues cannot be discussed due to limited page space.  The main
objective here is illustrative.

\subsection{Data}

In preparation for data acquisition by the Planck mission, the ability to perform
component separation is evaluated by resorting to a realistic set of simulated data which
is developed within the Planck collaboration as the `Planck sky model' (\textsc{PSM}).
This suite of programs is used to generate random realizations of the sky at the Planck
frequency range compatible with the present knowledge of all identified component
emissions and aims at capturing many of the intricacies (both from the sky and from the
instrument) expected from real data.

\begin{figure}[f]
  \newcommand\macol[2]{
    \begin{array}{c}
      \includegraphics[width=2cm]{\thumbdir/map--2048-#1planck-030--.jpg}\\
      \includegraphics[width=2cm]{\thumbdir/map--2048-#1planck-070--.jpg}\\
      \includegraphics[width=2cm]{\thumbdir/map--2048-#1planck-217--.jpg}\\
      \includegraphics[width=2cm]{\thumbdir/map--2048-#1planck-857--.jpg}\\
      \mbox{\tiny{#2}}
    \end{array} 
  }
  \centering
  \begin{displaymath} 
    \macol{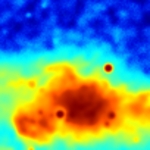}{data} = 
    \macol{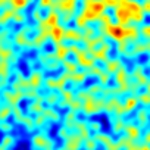}{CMB} + 
    \macol{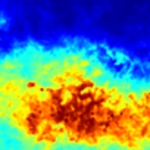}{Galactic} + 
    \macol{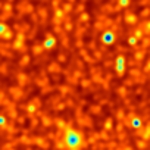}{SZ} + 
    \macol{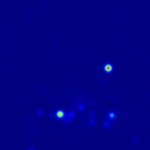}{PS} + 
    \macol{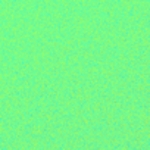}{noise}
  \end{displaymath}
  \caption{A zoom on a patch of size 17$\times$17 degrees, close to
    the Galactic plane for the channels at 30, 70, 217 and 857 GHz.
    The figure shows (in arbitrary color scales) the contributions of
    each physical component.}
  \label{fig:decinputs}
\end{figure}

We use sky maps simulated at the 9 frequencies of Planck channels.  The instrument point
spread function is modelled as a Gaussian beam whose width decreases with increasing
frequency: the beams' FWHM are 33, 24, 14, 10, 7.1, 5, 5, 5, 5 in arcminutes.  The beam
effect is visible in figure \ref{fig:decinputs} where the angular size of the galaxy
clusters (via the SZ effect), point sources and \textsc{CMB} anisotropies decreases from
top to bottom reflecting the increasing resolution.

\begin{figure}[f]
  \centering
  \includegraphics[width=7cm]{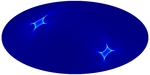}%
  \includegraphics[width=7cm]{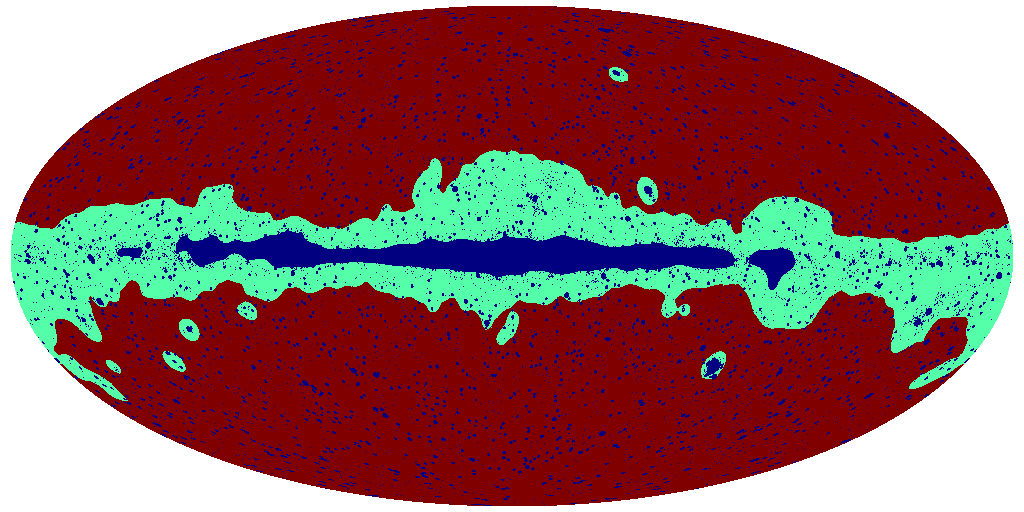}    
  \caption{ Left: Hit-counts map for the 30 GHz channel.  In each pixel, noise variance is
    inversely proportional to it. The number of hits in this map goes from around 40 to 2320.
    Right: the zones (masks) used in the analysis: polar zone (red), intermediate zone (green), Galactic
    zone and point source masking (blue) which is ignore in our analysis.}
  \label{fig:hitcount+mask}
\end{figure}

Planck does not take a `snapshot' of the sky.  Rather, sky maps are (painfully) computed
from sky scans.  Due to various technological constraints, the scanning strategy does
\emph{not} guarantee that all pixels are seen (or `hit') equally often.  As an important
consequence, the variance of the noise in each pixel depends on its hit-count.  In our
simulated data set, the noise is modelled as Gaussian, independent from channel to
channel, from pixel to pixel and with a variance inversely proportional to the hit-count.
See fig.~\ref{fig:hitcount+mask} for a sky map of hit counts corresponding to a one year
survey. 
The noise variance in one pixel for one hit depend on the frequency channel and are given
for the 9 channels by 1027, 1434, 2383, 1245, 753.6, 609.1, 424.5, 154.9, 71.8
$\mu\mathrm{K}^2$ for the maps used in this simulation.

\begin{figure}[f]
  \centering
  \includegraphics[width=5cm]{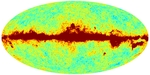}
  \includegraphics[width=5cm]{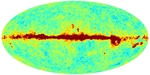}
  \includegraphics[width=5cm]{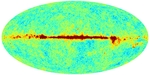}\\
  \includegraphics[width=5cm]{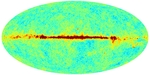}
  \includegraphics[width=5cm]{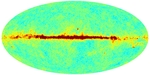}
  \includegraphics[width=5cm]{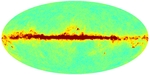}\\
  \includegraphics[width=5cm]{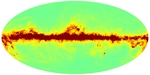}
  \includegraphics[width=5cm]{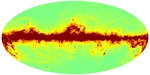}
  \includegraphics[width=5cm]{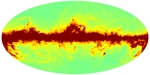}\\
  \caption{Simulated full-sky observations at the Planck frequencies:
    30, 44, 70, 100, 143, 217, 353, 545, 847 GHz.  Color scale: $\pm
    0.5$ mKRJ}.
  \label{fig:inputs}
\end{figure}

The full-sky maps used as inputs to our experiments are shown on figure~\ref{fig:inputs}.
Figure~\ref{fig:decinputs} zooms in and shows the various physical components used as
ingredients: \textsc{CMB}, Galactic emissions, galaxy clusters (via the SZ effect),
point source emissions (due to infrared- and radio-galaxies).  Galactic emission is shown
as a single component but is actually made up of three physical components due to
free-free, synchrotron and dust emissions.

\subsection{Building localized statistics}

The data are processed in Fourier space.  On the sphere, a Fourier basis for band limited
functions is a doubly-indexed set $\left\{ Y_{\ell m}(\xi) \ |\
  0\leq\ell\leq\ell_\mathrm{max}, \ -\ell\leq m\leq\ell \ \right\}$ of orthonormal
functions with the property that $\Delta Y_{\ell m} = -\ell(\ell+1)Y_{\ell m}$ where
$\Delta$ is the spherical Laplacian.  Index $\ell$ is called the angular frequency.
The coordinates $a_\mathrm{\ell m}$ of a spherical function on such a basis are called its
`spherical harmonic coefficients'.  The quantity $\widehat c_\ell =
\frac1{2\ell+1}\sum_{m=-\ell}^{m=+\ell}|a_{\ell m}|^2$ is the angular spectrum of the
function.  If this function is a realization of a stationary process, then $c_\ell =
\E\widehat c_\ell$ is the angular spectrum of the process.
The \textsc{CMB} is thought to be a realization of Gaussian stationary process; accurate
estimation of its angular spectrum is one the main scientific products expected from
\textsc{CMB} observations.

We compute localized statistics in both space and frequency: first the sphere is
decomposed in three zones: a small zone where Galactic emission is so strong that no
processing is attempted there, a second zone farther away from the Galactic center and
finally a zone of weak Galactic emission which includes both ecliptic poles.  The
corresponding masks are shown on fig~\ref{fig:hitcount+mask}.
We also localize in Fourier space by defining $Q=120$ domains which correspond to $120$
bins of angular frequency.  The width $\delta_\ell$ of these spectral bins increases with
the angular frequency $\ell$, as follows: 
$\delta_\ell=2$ for $0\leq\ell\leq29$, %
$\delta_\ell=5$ for $30\leq\ell\leq149$, %
$\delta_\ell=10$ for $150\leq\ell\leq419$, %
$\delta_\ell=20$ for $420\leq\ell\leq1199$, %
$\delta_\ell=50$ for $1200\leq\ell\leq2000$. %


For each of the two spatial domains, we build spectral statistics as follows: 
the zone of interest is isolated by using (an apodized version of) the masks shown on
fig~\ref{fig:hitcount+mask}; 
the spherical harmonic coefficients for each of the 9 frequency channels are computed and
collected in a $9\times 1$ vector $X_{lm}$ for each value of the pair $(\ell,m)$ up to
$\ell_\mathrm{max}=2000$; 
spectral matrices are then computed for each $\ell\leq 2000$ as $\widehat \bR_\ell =
\frac1{2\ell+1}\sum_m X_{\ell m} X_{\ell m}\adj$.  
At this stage, since symmetric beams are assumed, their effect can be corrected as
$\widehat \bR_\ell \leftarrow W_\ell \widehat \bR_\ell W_\ell$ where $W_\ell$ is a
diagonal matrix.
We also correct for incomplete sky coverage by dividing each $\R_\ell$ by a factor
$f_\mathrm{sky}$ which is the fraction of the sky left after masking.
Finally, the spectral matrices are binned into 
$\widehat \bR_q = \frac{\sum_{\ell} h_q(\ell) (2\ell+1) \widehat\bR_\ell}{\sum_{\ell}
  h_q(\ell) (2\ell+1)}$ where $h_q(\ell)$ is the top hat function for the $q$th bin.
These matrices collect an effective number $p_q$ of Fourier modes given by $p_q =
f_\mathrm{sky} \sum_{\ell} h_q(\ell) (2\ell+1)$.

\subsection{Component analysis}

We now fit a parametric model to the localized covariance matrices built at previous
section.  We fit independently the two zones of interest and then stitch the resulting
maps together.  In this illustrative example, we fit a model made of $C=4$ components
\begin{displaymath}
  \bR_q = \bR_q^\text{cmb} +\bR_q^\text{sz} +\bR_q^\text{gal} +\bR_q^\text{noise} 
  \qquad 1\leq q \leq 120 
\end{displaymath}
with the following parametrization/constraints:
\begin{compactitem}
\item The CMB component has a known emission law $\ba_\text{cmb}$ but unknown angular
  spectrum $c(q)$ so $\bR_q^\text{cmb}= \ba_\text{cmb} \ba_\text{cmb}\adj c_\text{cmb}(q)$
  and $\theta^\text{cmb} = \bigl\{ c_\text{cmb} (q)\bigl\}$.
  
\item The SZ component, as the CMB, is fully coherent and has a known emission law.
  Hence, we take $\bR_q^\text{sz}= \ba_\text{sz} \ba_\text{sz}\adj c_\text{sz}(q)$ and
  $\theta^\text{sz} = \bigl\{ c_\text{sz} (q)\bigl\}$.

\item We try to capture all Galactic emission (which is far from being coherent) 
  {\emph{together with emission from other galaxies -- the `point source' component}}, in a
  4-dimensional component without other constraints, that is, we set up the `Galactic'
  component model as $\bR_q^\text{gal} = \bA_\text{gal} \bP_q \bA_\text{gal}\adj$ with
  $\bA_\text{gal}$ an unconstrained $9\times4$ matrix and $\bP_q$ a $4\times4$ an
  unconstrained positive matrix so $ \theta^\text{gal} = \bigl\{ \bA_\text{gal} , \bP_q
  \bigl\}$. 
  
\item For the noise contribution, we rely almost entirely on the instrument
  characterization: only $m$ global parameters are adjustable as described in the third
  example of component in section \ref{sec:mylib}, that is we set up: $\bR_q^\text{noise}
  = \diag (\alpha_1n_{1q}^2, ..., \alpha_mn_{mq}^2)$ where $n_{iq}$ is known so
  $\theta^\text{noise}=\{\alpha_i\}$.  The fixed spectra $n_{iq}$ are computed from the
  hit-counts maps and other figures given above.

\end{compactitem}
Hence the global parameter vector is $\theta=[ c_\text{cmb}(q), c_\text{sz}(q),
\bA_\text{gal}, \bP_q , \alpha_i]$.

Note that the angular power spectrum of the \textsc{cmb} is obtained as a by-product of
the global fitting procedure.


\subsection{Some results}

We show and comment some results from our best-fit model.

\begin{figure}[f]
  \centering
  \includegraphics[width=12cm]{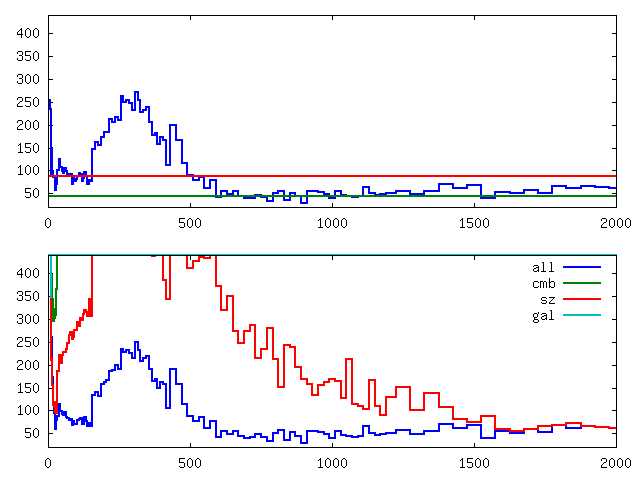}
  \caption{
    Top: Spectral mismatch versus angular frequency for the best-fit model.
    Bottom: mismatch when any one of the components (but noise) is removed.
    See text for more details.
  }
  \label{fig:mismatch}
\end{figure}

\textbf{Goodness of fit}.  
The top panel of figure~\ref{fig:mismatch} displays the values of $ w_q K(\widehat\R_q,
\R_q(\hat\theta))$ versus the spectral domain index $q$ (actually: versus the average
angular frequency $\ell$ for this domain).  The overall mismatch measure $\phi(\theta)$ is
just the sum of these quantities over all domains: see eq.~(\ref{eq:idcrit}).
It is possible to predict the average mismatch value (in the asymptotic regime of many
Fourier modes per domain) when the model holds.  This provides at least a reference value
in terms of goodness of fit.  This value (and twice this value) are displayed as a
horizontal line on fig.~\ref{fig:mismatch}.  The fit appears good for $\ell\geq 500$ but
not so good for $200\leq\ell\leq500$.  This reflects the difficulty of modelling the
complex galactic structure (since this component is important at this range scale) and
also possibly the fact that the assumptions required for predicting the average mismatch are
not met in this domain.  The bottom panel of the figure shows how the mismatch increases
when the contribution of any one of the components (except noise) is removed from the
model.  In this case, the mismatch explodes, showing that all fitted components are
significant here.

\begin{figure}[f]
  \centering
  \newcommand\dispauto[1]{\includegraphics[width=7cm]{\newfigdir/auto_#1.png}}
  \dispauto{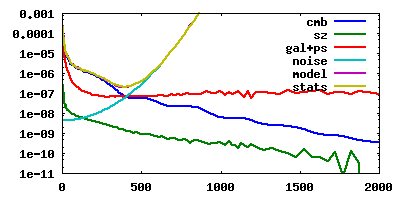}\dispauto{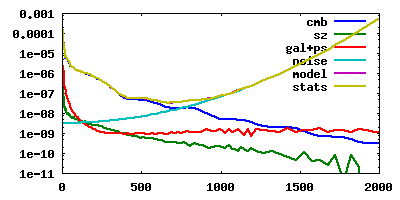}\\
  \dispauto{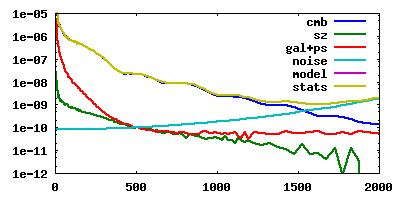}\dispauto{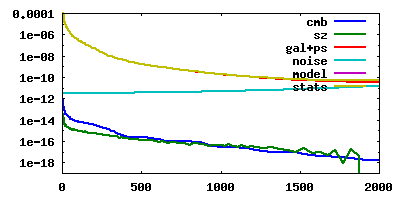}
  \caption{Angular spectra on channels at 30, 70, 217, 857 GHz
    estimated with our best in the polar zone (mKRJ$^2$).  
    The line labelled \textit{stats} shows the values of
    $[\widehat\R_q]_{jj}$; the line labelled \textit{model} the values
    of $[\R_q(\hat\theta)]_{jj}$; the latter is hidden by the former
    almost everywhere.}
  \label{fig:specsmic}
\end{figure}

\textbf{Power decomposition at sensor level.}  
The decomposition (\ref{eq:decRq}) allows to find the contribution of
each component on each detector by domain, \textit{i.e.} as a function
of angular frequency in the present case.  Figure~\ref{fig:specsmic}
displays for all detectors ($j=1,\ldots 9$) the values of
$[\R_q^c(\hat\theta)]_{jj}$ versus the center frequency of the $q$th
spectral domain.  In other words it shows estimated band-averaged
angular spectra for all fitted components.  It also shows the sum of
these contributions over all components as well as the values of
$[\widehat\R_q]_{jj}$.  Actually the plot of the former quantity is
almost entierly masked by the plot of the latter, evidencing a near
perfect fit of the global angular spectrum.

\begin{figure}[f]
  \centering
  \newcommand\dispcross[1]{\includegraphics[width=7cm]{\newfigdir/cross_#1.png}}
  \dispcross{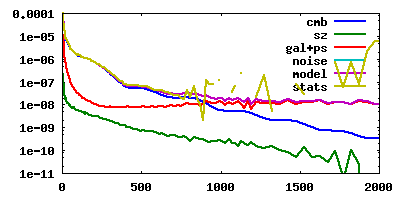}\dispcross{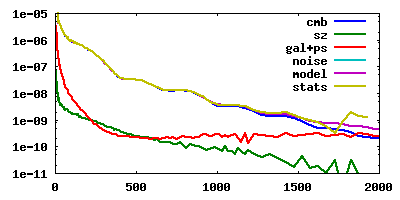}\\
  \dispcross{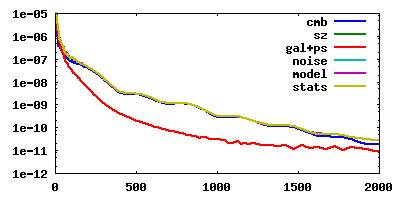}\dispcross{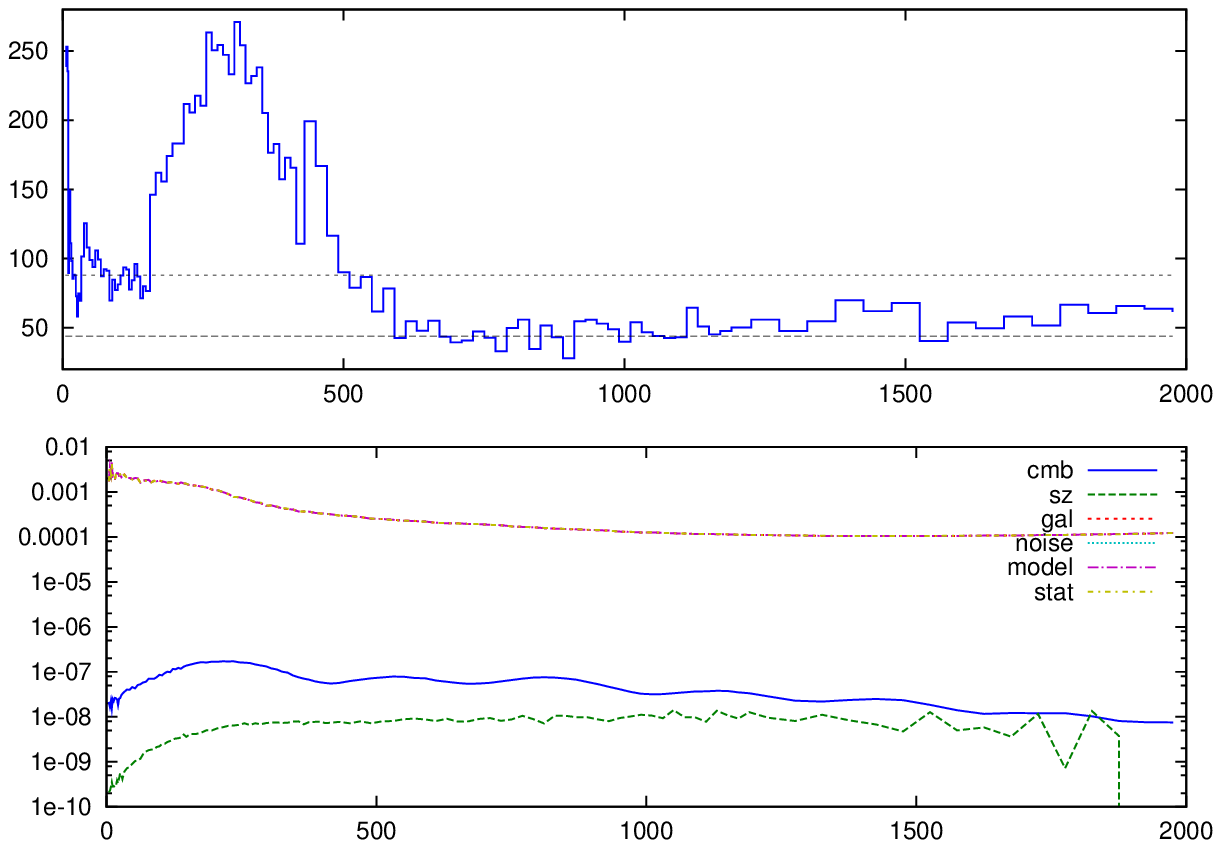}
  \caption{Cross-spectra for the channel pairs 30-70 GHz, 70-143 GHz, 143-353 GHz and
    353-857 GHz and our fit with cmb+sz+(Galaxy+PS)+noise in polar zone (mKRJ$^2$). Same
    conventions as in figure \ref{fig:specsmic}.}
  \label{fig:specsmiccross}
\end{figure}

Figure~\ref{fig:specsmiccross} shows the cross-spectra between
channels, that is the values of $[\R_q^c(\hat\theta)]_{ij}$ for $4$
selected pairs of channels. 

\begin{figure}[f]
  \centering
  \includegraphics[height=6cm]{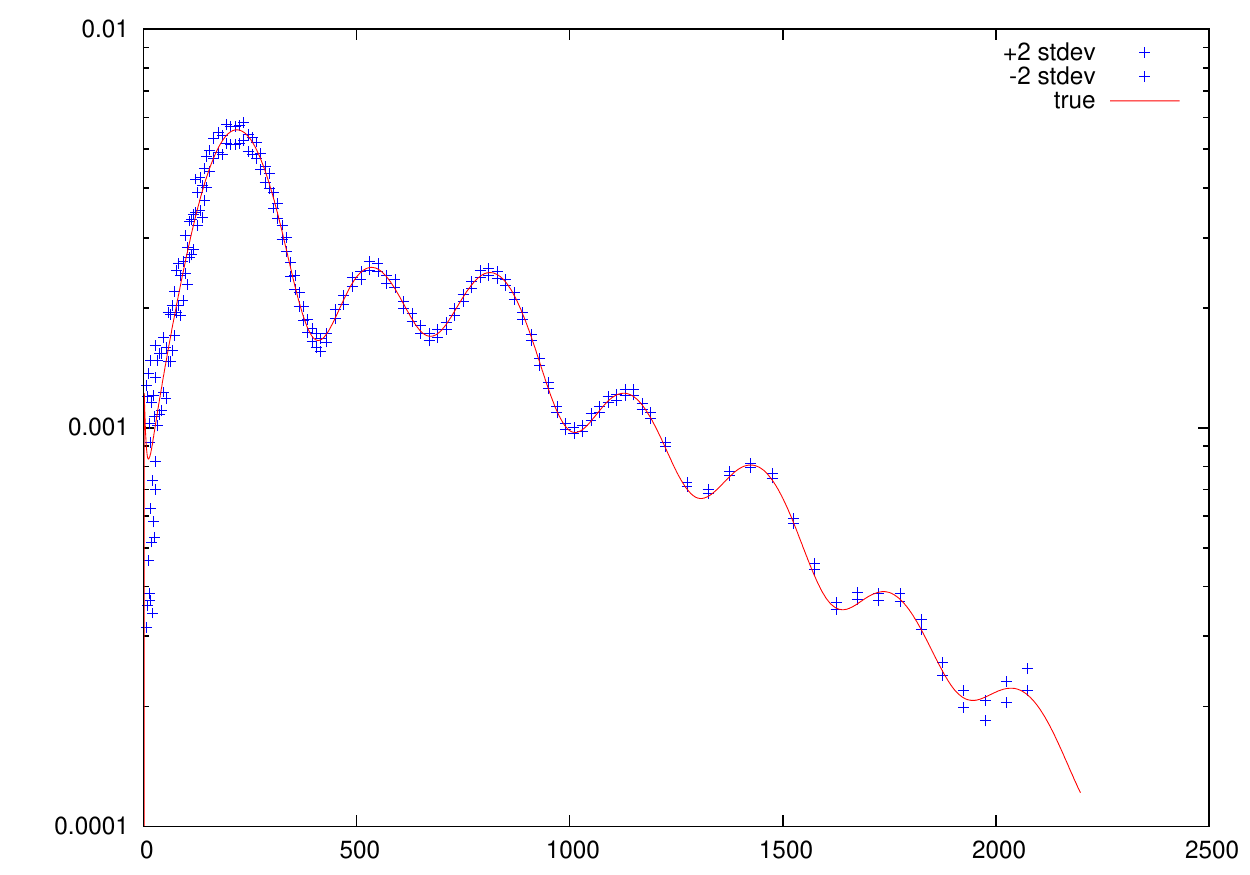}%
  \includegraphics[height=6cm]{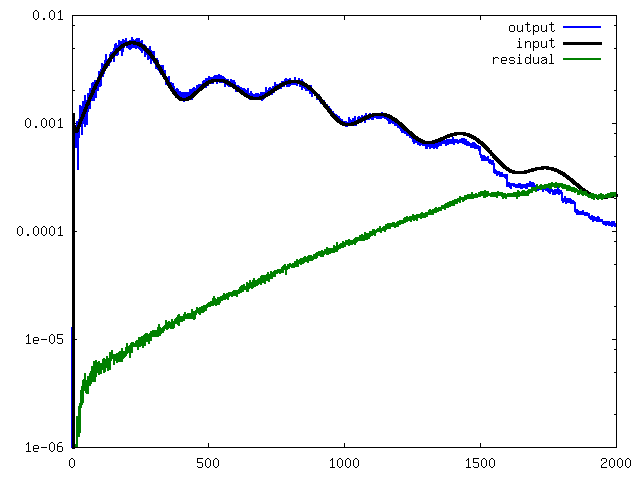}
  \caption{ 
    Left: Estimated angular spectrum of the \textsc{CMB} (in polar zone) in mKCMB$^2$ with
    $\pm 2\sigma$ error bars given by the Fisher information matrix.  We (conventionally)
    plot $\ell(\ell+1)c_{\ell}/2\pi$ rather than the plain angular spectrum $c_{\ell}$.  
    Right: Angular spectrum of the reconstructed \textsc{CMB} and of the residual in
    mKCMB$^2$. The red and the black curves are the same reference input \textsc{CMB} spectrum
    which has been used to generate the \textsc{CMB} component.}
  \label{fig:cl_est+res}
\end{figure}

\textbf{CMB angular spectra.}  
Left panel of figure~\ref{fig:cl_est+res} shows the estimated angular spectrum $\hat
c_\ell^\mathrm{cmb}$ with error bars computed from the Fisher information matrix.
The angular spectrum of the CMB process is known in this simulated data set and displayed
as a solid line.  An excellent recovery is observed up to angular frequency $\ell = 2000$,
together with plausible error bars.

\begin{figure}[f]
  \newcommand\cmbmap[1]{\includegraphics{\newfigdir/#1.jpg}}
  \centering
  \cmbmap{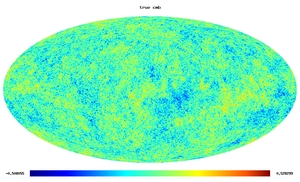}
  \cmbmap{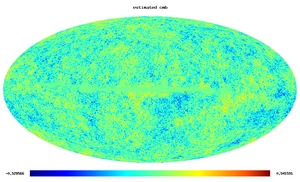}  
  \\
  \cmbmap{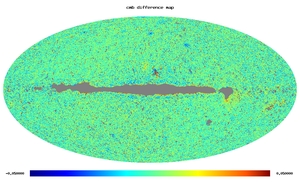}  
  \caption{Top: input CMB map; middle: estimated map outside of a small Galactic mask (color
    scale: $\pm 0.5$~mK~CMB).  Bottom: difference map (color scale: $\pm 0.05$~mK~CMB).}
  \label{fig:reccmb}
\end{figure}

\textbf{CMB map}.
The CMB map recovered through the localized Wiener filter~(\ref{eq:locwiener}) is shown in
Figure~\ref{fig:reccmb}.  It has no visible contamination but reconstruction is not over
the whole sky: the small Galactic region (shown on fig.~\ref{fig:hitcount+mask}) is not
reconstructed because the Galactic emission is so strong that avoiding any contamination
is very difficult.
The bottom map of fig.~\ref{fig:reccmb} is the residual \textit{i.e.} the error between
true and reconstructed CMB maps.  Some Galactic residuals are now visible indicating that
the localization strategy should be improved (the Galactic emission is known to vary a lot
across the sky, and especially near the Galactic plane).

The angular spectrum of the \emph{residual} map (CMB error map) is shown on the right
panel of fig.~\ref{fig:cl_est+res}.  The linear (on this log scale) trend indicates that
the error is noise-dominated (as also seen from the residual map).


\begin{figure}[f]
  \newcommand\sumgal[1]{\includegraphics{\newfigdir/map-2048-#1.jpg}}

  \centering
  \sumgal{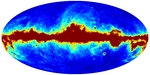}\sumgal{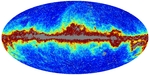}\sumgal{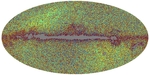}\\
  \sumgal{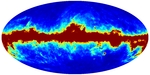}\sumgal{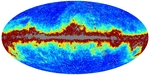}\sumgal{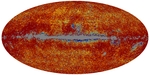}\\
  \sumgal{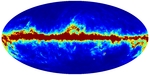}\sumgal{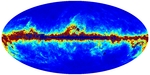}\sumgal{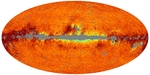}\\
  \sumgal{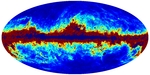}\sumgal{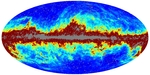}\sumgal{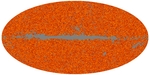}\\
  \caption{The Galactic component seen by the 30, 70, 217 and 857 GHz channels. 
    Left column: dust + free-free + synchrotron from the input maps in MJ/sr. Scale between 0 and 0.01 for the 30 and 70 GHz channels, between 0 and 0.1 for the 217 GHz channel and between 0 and 7 for the 857 GHz channel;
    center column: the 4-dimensional component found by SMICA with the same scale range;
    right column: the difference, rescaled by a factor of 10 to emphasize the errors.}
  \label{fig:galrec}
\end{figure}

\textbf{Galactic and point source emission reconstruction}.
The most complex component is due to the radiation from the interstellar medium in our
Galaxy (Galactic emission) as well as other galaxies (which contribute most of the
emission of extragalactic `point sources').  The present work focuses
on CMB reconstruction, and no attempt has been made to separate the Galactic
emissions on the basis of the astrophysical radiation emission process (synchrotron
emission, `free-free' emission, or grey body emission from dust particles), nor on the
basis of their origin (within our own Galaxy, or in other galaxies).  For this reason, the
sum of these emissions {\emph{outside our Galactic + point source mask}} has been
tentatively represented by a 4-dimensional `catch all' model as explained above. 
The result of the separation
for this component is shown in fig.~\ref{fig:galrec}. In spite of small discrepancies in
the reconstruction on large angular scales, the result is quite satisfactory. Angular
power spectra for this model, displayed in fig.~\ref{fig:specsmic} clearly correspond to
the sum of the steep Galactic power spectrum, and a flat plateau at large angular
frequency $\ell$, due to the contribution of emissions from the large number of
extragalactic sources not masked by our point source mask.

\sautpage
\section{Conclusion}

The component separation technique discussed in this paper offers great modeling
flexibility from the realization that separation can start with \emph{covariance matrix
  separation} ---\emph{i.e.} the identification of individual component terms in the
domain-wise decomposition~(\ref{eq:decRq})--- followed by \emph{data separation} according
to~(\ref{eq:wiener}).
Too much flexibility may also introduce difficulties, though.  Indeed, whether or not
minimizing the covariance matching criterion $\phi(\theta)$ leads to \emph{uniquely}
identified components depends on the particular choice of component models.  In our
approach, however, the amount of constraints imposed on any component is fully tunable.
By using more or less constrained components, the method ranges from totally blind to
semi-blind, to non-blind.

Some other good points are the following.  
\textbf{Speed}: the method is potentially fast because large data sets are compressed into
a much smaller set of covariance matrices. 
\textbf{Accuracy}: the method is potentially accurate because it can model complex
components and then recover separated data via local Wiener filters which are naturally
adapted to the local SNR conditions. 
\textbf{Noise}: the method can take noise into account without increased complexity since
noise is not processed differently from any other component.  
\textbf{Prior}: the implementation also allows for easy inclusion of prior information
about a component $c$ if it can be cast in the form of a prior probability distribution
$p_c(\theta^c)$ in which case one only need to subtracting $\log p_c(\theta^c)$ from
$\phi(\theta)$ and the related changes can be delegated to the component code. 
\textbf{Varying resolution}: in our application, and possibly others, the input channels
are acquired by sensors with channel-dependent resolution.  Accurate component separation
requires to take this effect into account.  This can be achieved with relative simplicity
if the data coefficients entering in $\widehat\R_q$ are Fourier (or spherical harmonic)
coefficients. 
\textbf{Built-in goodness of fit} via the mismatch measure~(\ref{eq:idcrit}).

This paper combines several ideas already known in the ICA literature: lumping together
correlated components into a single multidimensional component is in \cite{MICA:ICASSP};
minimization of a covariance-matching contrast $\phi(\theta)$ derived from the
log-likelihood of a simple Gaussian model is found for instance in~\cite{PhamGauss}; the
extension to noisy models is already explained in~\cite{mnras:MCMD}.  The current paper
goes one step further by showing how arbitrarily structured components can be separated
and how the related complexity can be managed at the software level by a library of
components.

\textbf{About Gaussianity}.  The specific choice of the matching criterion
(\ref{eq:idcrit}) stems from a Gaussian model for the signal coefficients and the linear
Wiener filter~(\ref{eq:wiener}) is optimal only for Gaussian signals.  Hence, there is no
doubt that an improved statistical efficiency could be gained by resorting to non Gaussian
models and to non-linear filtering for strongly non Gaussian data.  However, this is more
easily said than done, non Gaussian modeling/processing being often more difficult and
costly to implement.  By sticking to simple Gaussian assumptions, we can afford to model a
non trivial correlation structure (through domains and through sensors) so it is not clear
yet what the good trade-off is.  It may depend very much on the scientific objectives
(recovery of CMB versus recovery of other components, ability to predict estimation
errors, \ldots) and technical constraints (\textit{e.g.} fast codes are important for
assessment via Monte-Carlo runs).
Intensive work is in progress within the Planck collaboration to assess the performance of
various approaches for CMB analysis.

\section*{Acknowledgments}

We acknowledge the use of Healpix~\cite{healpix} for all sphere-based computations.  Our
code is implemented in Octave (\texttt{octave.org}).  Synthetic data are made available
thanks to the Planck collaboration. Maude Martin was partially supported by Astromap and
Cosmostat grants, two ACI programs of CNRS.


\begin{thebibliography}{10}

\bibitem{plancksite}
The {P}lanck mission of {ESA}.

\bibitem{2003ApJS..148...97B}
C.~L. {Bennett}, R.~S. {Hill}, G.~{Hinshaw}, M.~R. {Nolta}, N.~{Odegard},
  L.~{Page}, D.~N. {Spergel}, J.~L. {Weiland}, E.~L. {Wright}, M.~{Halpern},
  N.~{Jarosik}, A.~{Kogut}, M.~{Limon}, S.~S. {Meyer}, G.~S. {Tucker}, and
  E.~{Wollack}.
\newblock {First-Year Wilkinson Microwave Anisotropy Probe (WMAP) Observations:
  Foreground Emission}.
\newblock {\em apjs}, 148:97--117, September 2003.

\bibitem{2006MNRAS.373..271B}
A.~{Bonaldi}, L.~{Bedini}, E.~{Salerno}, C.~{Baccigalupi}, and G.~{de Zotti}.
\newblock {Estimating the spectral indices of correlated astrophysical
  foregrounds by a second-order statistical approach}.
\newblock {\em mnras}, 373:271--279, November 2006.

\bibitem{1999NewA....4..443B}
F.~R. {Bouchet} and R.~{Gispert}.
\newblock {Foregrounds and CMB experiments I. Semi-analytical estimates of
  contamination}.
\newblock {\em New Astronomy}, 4:443--479, September 1999.

\bibitem{MICA:ICASSP}
Jean-Fran\c{c}ois Cardoso.
\newblock Multidimensional independent component analysis.
\newblock In {\em Proc. ICASSP '98. Seattle}, pages 1941--1944, 1998.

\bibitem{2003MNRAS.346.1089D}
J.~{Delabrouille}, J.-F. {Cardoso}, and G.~{Patanchon}.
\newblock {Multidetector multicomponent spectral matching and applications for
  cosmic microwave background data analysis}.
\newblock {\em mnras}, 346:1089--1102, December 2003.

\bibitem{LNP-Valencia}
Jacques Delabrouille and Jean-Fran\c{c}ois Cardoso.
\newblock {\em Data Analysis in Cosmology}, chapter Diffuse source separation
  in {CMB} observations.
\newblock {L}ecture {N}otes in {P}hysics. Springer, 2007.
\newblock Editors: Vicent J. Martinez, Enn Saar, Enrique Martinez-Gonzalez,
  Maria Jesus Pons-Borderia.

\bibitem{mnras:MCMD}
Jacques Delabrouille, Jean-Fran\c{c}ois Cardoso, and Guillaume Patanchon.
\newblock Multi--detector multi--component spectral matching and applications
  for {CMB} data analysis.
\newblock {\em Monthly Notices of the Royal Astronomical Society},
  346(4):1089--1102, December 2003.
\newblock also available as http://arXiv.org/abs/astro-ph/0211504.

\bibitem{2006NewAR..50..861E}
H.~K. {Eriksen}, C.~{Dickinson}, C.~R. {Lawrence}, C.~{Baccigalupi}, A.~J.
  {Banday}, K.~M. {G{\'o}rski}, F.~K. {Hansen}, E.~{Pierpaoli}, and M.~D.
  {Seiffert}.
\newblock {Bayesian foreground analysis with CMB data}.
\newblock {\em New Astronomy Review}, 50:861--867, December 2006.

\bibitem{healpix}
K.~M. Gorski et~al.
\newblock Healpix -- a framework for high resolution discretization, and fast
  analysis of data distributed on the sphere.
\newblock {\em Astrophys. J.}, 622:759--771, 2005.

\bibitem{2006ApJ...648..784H}
F.~K. {Hansen}, A.~J. {Banday}, H.~K. {Eriksen}, K.~M. {G{\'o}rski}, and P.~B.
  {Lilje}.
\newblock {Foreground Subtraction of Cosmic Microwave Background Maps Using
  WI-FIT (Wavelet-Based High-Resolution Fitting of Internal Templates)}.
\newblock {\em apj}, 648:784--796, September 2006.

\bibitem{1998MNRAS.300....1H}
M.~P. {Hobson}, A.~W. {Jones}, A.~N. {Lasenby}, and F.~R. {Bouchet}.
\newblock {Foreground separation methods for satellite observations of the
  cosmic microwave background}.
\newblock {\em mnras}, 300:1--29, October 1998.

\bibitem{2002MNRAS.334...53M}
D.~{Maino}, A.~{Farusi}, C.~{Baccigalupi}, F.~{Perrotta}, A.~J. {Banday},
  L.~{Bedini}, C.~{Burigana}, G.~{De Zotti}, K.~M. {G{\'o}rski}, and
  E.~{Salerno}.
\newblock {All-sky astrophysical component separation with Fast Independent
  Component Analysis (FASTICA)}.
\newblock {\em mnras}, 334:53--68, July 2002.

\bibitem{2003MNRAS.345.1101M}
E.~{Mart{\'{\i}}nez-Gonz{\'a}lez}, J.~M. {Diego}, P.~{Vielva}, and J.~{Silk}.
\newblock {Cosmic microwave background power spectrum estimation and map
  reconstruction with the expectation-maximization algorithm}.
\newblock {\em mnras}, 345:1101--1109, November 2003.

\bibitem{ParraSpence:ieee}
Lucas Parra and Clay Spence.
\newblock Convolutive blind source separation of non-stationary sources.
\newblock {\em IEEE Trans. on Speech and Audio Processing}, pages 320--327, may
  2000.

\bibitem{nstatIEEESP}
Dinh-Tuan Pham and Jean-Fran\c{c}ois Cardoso.
\newblock Blind separation of instantaneous mixtures of non stationary sources.
\newblock {\em IEEE Trans. on Sig. Proc.}, 49(9):1837--1848, September 2001.

\bibitem{PhamGauss}
D.T. Pham.
\newblock Blind separation of instantaneous mixture of sources via the
  {G}aussian mutual information criterion.
\newblock {\em Signal Processing}, (4):855--870, 2001.

\bibitem{2002AIPC..617..125S}
H.~{Snoussi}, G.~{Patanchon}, J.~F. {Mac{\'{\i}}as-P{\'e}rez},
  A.~{Mohammad-Djafari}, and J.~{Delabrouille}.
\newblock {Bayesian blind component separation for cosmic microwave background
  observations}.
\newblock In R.~L. {Fry}, editor, {\em Bayesian Inference and Maximum Entropy
  Methods in Science and Engineering}, volume 617 of {\em American Institute of
  Physics Conference Series}, pages 125--140, May 2002.

\bibitem{2002MNRAS.336...97S}
V.~{Stolyarov}, M.~P. {Hobson}, M.~A.~J. {Ashdown}, and A.~N. {Lasenby}.
\newblock {All-sky component separation for the Planck mission}.
\newblock {\em mnras}, 336:97--111, October 2002.

\bibitem{1996MNRAS.281.1297T}
M.~{Tegmark} and G.~{Efstathiou}.
\newblock {A method for subtracting foregrounds from multifrequency CMB sky
  maps}.
\newblock {\em mnras}, 281:1297--1314, August 1996.

\bibitem{2000ApJ...530..133T}
M.~{Tegmark}, D.~J. {Eisenstein}, W.~{Hu}, and A.~{de Oliveira-Costa}.
\newblock {Foregrounds and Forecasts for the Cosmic Microwave Background}.
\newblock {\em apj}, 530:133--165, February 2000.

\end{thebibliography}

\end{document}